\documentclass[preprint,11pt,3p]{elsarticle}

\usepackage{amssymb}
\usepackage{amsmath}
\usepackage{listings} % Used for writing codes in latex files
\usepackage{showexpl} % Shows examples of latex along the code
\usepackage{color} % Used to define color in RGB format
\usepackage{amsthm} % Used for theorems/lemmas
\usepackage{alltt}
\usepackage[export]{adjustbox} % to adjust figure to left/right/center

\usepackage[ruled,vlined]{algorithm2e}
\usepackage{algpseudocode}
\usepackage{amssymb}
\usepackage{amsmath}
\usepackage{amsfonts}
\usepackage{booktabs}
\usepackage{enumerate}
\usepackage{fancyhdr}
\usepackage{float}
\usepackage{graphicx}
\usepackage{lastpage}
\usepackage{mathrsfs}
\usepackage{hyperref}
\hypersetup{
    colorlinks=true,
    linkcolor=blue,
    filecolor=magenta,      
    urlcolor=cyan,
}

\usepackage{appendix}
\usepackage{tikz}
\usetikzlibrary{decorations.text}
\usetikzlibrary{shapes,arrows}
\usepackage{pgfplots}
\usepackage{pgfplotstable}
\pgfplotsset{compat=newest}

\usetikzlibrary{calc}
\usetikzlibrary{spy}
\usetikzlibrary{positioning}
\usepackage{pdftexcmds}
\usetikzlibrary{external}
\usetikzlibrary{positioning}
\usetikzlibrary{arrows}
\usetikzlibrary{decorations.markings}
\usetikzlibrary{decorations.text}
\usetikzlibrary{backgrounds}
\usetikzlibrary{spy}
\usetikzlibrary{calc,patterns,decorations.pathmorphing,decorations.markings}
\usetikzlibrary{decorations.pathreplacing}
\usepgfplotslibrary{patchplots}

\usepackage{chngcntr}
\usepackage{etoolbox}
\usepackage{lipsum}
\usepackage{bm}

\makeatletter
\providecommand*{\input@path}{}
\g@addto@macro\input@path{{./}}% append
\makeatother

\makeatletter

\def\pgfplotstableread@openfile{%
    \def\pgfplotstable@loc@TMPa{\pgfutil@in@{ }}%
    \expandafter\pgfplotstable@loc@TMPa\expandafter{\pgfplotstableread@filename}%
    \ifpgfutil@in@
        \t@pgfplots@toka=\expandafter{\pgfplotstableread@filename}%
        \edef\pgfplotstableread@filename{\pgfplots@dquote\the\t@pgfplots@toka\pgfplots@dquote}%
    \fi
    \let\pgfplotstableread@old@crcr=\\%
    \def\\{\string\\}% just to make sure we don't try to open inline table data...
    \openin\r@pgfplots@reada=\csname pgfk@/pgfplots/table file path\endcsname\pgfplotstableread@filename.tex
    \ifeof\r@pgfplots@reada
        \openin\r@pgfplots@reada=\csname pgfk@/pgfplots/table file path\endcsname\pgfplotstableread@filename\relax
    \else
        \pgfplots@warning{%
            You requested to open table '\pgfplotstableread@filename', but there is also a '\pgfplotstableread@filename.tex'. 
            TeX will automatically append the suffix '.tex', so I will now open '\pgfplotstableread@filename.tex'.
            Please make sure you don't accidentally load TeX files - this may produce unrecoverable errors.}%
        \closein\r@pgfplots@reada
        \openin\r@pgfplots@reada=\pgfplotstableread@filename\relax
    \fi
    \ifeof\r@pgfplots@reada
        \pgfplotsthrow{no such table file}{\pgfplots@loc@TMPa}{\pgfplotstableread@filename}{Could not read table file '\csname pgfk@/pgfplots/table file path\endcsname\pgfplotstableread@filename'. In case you intended to provide inline data: maybe TeX screwed up your end-of-lines? Try `row sep=crcr' and terminate your lines with `\string\\' (refer to the pgfplotstable manual for details)}\pgfeov%
        \global\let\pgfplotstable@colnames@glob=\pgfplots@loc@TMPa
        \def\pgfplotstableread@ready{0}%
    \fi
    \pgfplots@logfileopen{\pgfplotstableread@filename}%
    \let\\=\pgfplotstableread@old@crcr
}

\makeatother

\pgfplotscreateplotcyclelist{blue to red}{%
  color=blue\\%
  color=red!10!blue\\%
  color=red!20!blue\\%
  color=red!30!blue\\%
  color=red!40!blue\\%
  color=red!50!blue\\%
  color=red!60!blue\\%
  color=red!70!blue\\%
  color=red!80!blue\\%
  color=red!90!blue\\%
  color=red\\%
}
\pgfplotscreateplotcyclelist{red to blue}{%
  color=red\\%
  color=red!90!blue\\%
  color=red!80!blue\\%
  color=red!70!blue\\%
  color=red!60!blue\\%
  color=red!50!blue\\%
  color=red!40!blue\\%
  color=red!30!blue\\%
  color=red!20!blue\\%
  color=red!10!blue\\%
  color=blue\\%
}

\pgfplotscreateplotcyclelist{red to blue fast}{%
  color=red\\%
  color=red!80!blue\\%
  color=red!60!blue\\%
  color=red!40!blue\\%
  color=red!20!blue\\%
  color=blue\\%
}

\pgfplotscreateplotcyclelist{2red to blue}{%
  color=red\\% 
  color=red\\%
  color=red!90!blue\\%
  color=red!90!blue\\%
  color=red!80!blue\\%
  color=red!80!blue\\%
  color=red!70!blue\\%
  color=red!70!blue\\%
  color=red!60!blue\\%
  color=red!60!blue\\%
  color=red!50!blue\\%
  color=red!50!blue\\%
  color=red!40!blue\\%
  color=red!40!blue\\%
  color=red!30!blue\\%
  color=red!30!blue\\%
  color=red!20!blue\\%
  color=red!20!blue\\%
  color=red!20!blue\\%
  color=red!10!blue\\%
  color=blue\\%
  color=blue\\%
}

\pgfplotsset{discard if/.style 2 args={x filter/.code={\ifnum\thisrow{#1}=#2\else\fi}}}

\usepackage{bm}

\newcommand{\lambdab}{{\boldsymbol \lambda}}

\newcommand{\thetab}{\boldsymbol \theta}

\newcommand{\Ab}{{\boldsymbol A}}

\newcommand{\Gb}{{\boldsymbol G}}
\newcommand{\Fb}{{\boldsymbol F}}

\newcommand{\Ib}{{\boldsymbol I}}

\newcommand{\Nb}{{\boldsymbol N}}
\newcommand{\Pb}{{\boldsymbol P}}

\newcommand{\Ub}{{\boldsymbol U}}
\newcommand{\Vb}{{\boldsymbol V}}
\newcommand{\Wb}{{\boldsymbol W}}

\newcommand{\fb}{{\boldsymbol f}}

\newcommand{\tb}{{\boldsymbol t}}
\newcommand{\ub}{{\boldsymbol u}}
\newcommand{\vb}{{\boldsymbol v}}
\newcommand{\wb}{{\boldsymbol w}}

\newcommand{\ee}{{ \bf e}}

\newcommand{\x}{{\boldsymbol x}}

\newcommand{\X}{{\boldsymbol X}}

\usepackage{amsmath}

\usepackage{nicefrac}
\usepackage{multirow}
\usepackage{hhline}
\usepackage{lineno}
\usepackage{import}
\usepackage{epstopdf}
\usepackage{subcaption}
\usepackage{mathtools}
\usepackage{transparent}

\usepackage{listings}
\usepackage{cancel}
\usepackage{empheq}
\usepackage{lineno}
\definecolor{tbf}{RGB}{255,0,0} % to be fixed
\definecolor{txue}{RGB}{0,0,255}

\journal{Computer Methods in Applied Mechanics and Engineering}

\begin{document}

\begin{frontmatter}

\author[label1]{Xinxin Wu}
%\author[label2]{Kaiqiang Sun}
%\author[label2]{Shaohua Yang}
%\author[label2]{Ye Xu}
\author[label1]{Yin Zhang}
\author[label1]{Sheng Mao}

\address[label1]{Department of Mechanics and Engineering Science, College of Engineering, Peking University, Beijing 100871, China}
%\address[label2]{School of Mechanical Engineering and Automation, Beihang University, 
%Beijing 100191, China}
%\address[label3]{College of Chemistry and Molecular Engineering, Peking University, 
%Beijing 100871, China}
% \title{Learning hyperelastic constitutive models from experimental data under partial differential equation constraints}
% \title{Discovering hyperelasticity models from experimental obsevations}
\title{Learning the physics-consistent material behavior from experimentally measurable data via PDE-constrained optimization}
%\title{A hybrid FEM-NN optimization method to learn the physics-constrained constitutive relations from full-field data}
%\author{Xinxin Wu$^a$, Kaiqiang Sun$^b$, Shaohua Yang$^b$, Ye Xu$^b$, Yin Zhang$^{a}$, Sheng Mao$^{a}$}
%\date{\normalsize a, Department of Mechanics and Engineering Science, College of Engineering, Peking University, Beijing 100871, P.R. China \\
%b, School of Mechanical Engineering and Automation, Beihang University, \\
%Beijing 100191, P.R. China}


\begin{abstract}
% \begin{linenumbers}
Constitutive models play a crucial role in materials science as they describe the behavior of the materials in mathematical forms.
Over the last few decades, the rapid development of manufacturing technologies have led to the discovery of many advanced materials with complex and novel behaviors, which in the meantime, have also posed great challenges for constructing accurate and reliable constitutive models of these materials.
In this work, we propose a data-driven approach to construct physics-consistent constitutive models for hyperelastic materials from experimentally measurable data, with the help of PDE-constrained optimization methods.
Specifically, our constitutive models are based on the physically augmented neural networks~(PANNs), which has been shown to ensure that the models are both physically consistent but also mathematically well-posed by construction.
Specimens with deliberately introduced inhomogeneity are used to generate the data, i.e., the full-field displacement data and the total external load, for training the model.
Using such approach, a considerably diverse pairs of stress-strain states can be explored with a limited number of simple experiments, such as uniaxial tension. 
A loss function is defined to measure the difference between the data and the model prediction, which is obtained by numerically solving the governing PDEs under the same geometry and loading conditions.
With the help of adjoint method, we can iteratively optimize the parameters of our NN-based constitutive models through gradient descent.
We test our method for a wide range of hyperelastic materials and in all cases, our methods are able to capture the constitutive model efficiently and accurately.
The trained models are also tested against unseen geometry and unseen loading conditions, exhibiting strong interpolation and extrapolation capabilities.
% \end{linenumbers}
\end{abstract}

\end{frontmatter}

\section{Introduction}
\label{Sec:Introduction}
Constitutive models are critically important in mechanics and material science, as they describe the mechanical behaviors of materials in mathematical forms.
In practical applications, accurate constitutive models are key to the prediction of the mechanical behaviors of materials, including elasticity, plasticity, fatigue, and fracture, playing an important role in areas such as civil engineering, structural design and optimization, biomedical engineering, etc.~\cite{belytschko2014nonlinear, bendsoe2013topology, eschenauer2001topology, humphrey2003continuum, fung2013biomechanics}.
In the last few decades, as with the advent of advanced manufacturing technology, materials with novel, sophisticated functionalities can now be readily fabricated. 
The complex physical behaviors of these materials makes it a challenge to construct reliable and predictive constitutive relations for them.
As a result, it becomes utterly important to develop efficient and accurate methods for the construction of complex material constitutive models.

Traditionally, the construction of constitutive relations often requires a model with undetermined material parameters proposed based on the observation of macroscopic material behavior~\cite{rivlin1948large, mooney1940theory, gent1996new} or microscopic material mechanisms~\cite{arruda1993three}.
The material parameters are calibrated by comparing the model prediction with data obtained from experiments such as uniaxial tension, where the deformation of the material can be approximately seen as uniform.
%Then, the material parameters are calibrated by the force-displacement data generated from simple loading experiments, e.g., uni/biaxial tension, pure shear and torsion tests.
However, how to choose the right model for calibration in the first place requires a lot of experience in material modeling.
Choosing an inappropriate model can lead to poor convergence in calibrating the material parameters and poor ability for further prediction.
Moreover, many existing models might not be able to capture the behavior of these new materials under complex loading conditions.
For example, for athermal fibrous networks, although the widely used eight-chain and micro-sphere constitutive models can accurately describe their axial behavior, these models can have poor performance in capturing their shear behavior~\cite{song2022hyperelastic}.
To do so, then it in turn requires more experimental characterization of these complex loading states, which can be hard to realize.
%When dealing with complex materials, the main shortcoming of traditional process can be summarized in two key points.
%First, a priori models often suffer from limited representational capability.
%For instance,  the wildly used eight chain and micro-sphere constitutive constitutive models, which can accurately describe the axial behavior of athermal fibrous networks have been shown to be inadequate in capturing the shear behavior\cite{song2022hyperelastic}.
%Second, the simple loading tests under a uniform strain state can only provide stress-strain information at a single point.
%In order to accurately describe the behavior of materials under various deformation states, a large number of experiments are needed to calibrate the material parameters\cite{ogden1997non, holzapfel2002nonlinear}, which can be costly and inefficient.

Fortunately, the recent developments in machine learning (ML) have provided us with a solution to construct a constitutive model that has strong representability using a data-driven approach and that is, to base our constitutive model on artificial neural networks~(ANNs) or neural networks (NNs) in short.
Artificial neural networks, as a universal approximator~\cite{hornik1991approximation}, is known to be extremely powerful in approximating high-dimensional functional relationship and have recently been successfully applied to model linear elasticity, elastoplasticity, hyperelasticity, damage models, etc.~\cite{shen2004neural, liu2020learning, liu2020neural, lu2020extraction, xu2021learning}
To ensure that the NN-based constitutive models are physically consistent, recent efforts have been devoted to guarantee that important physical principles, such as material frame-indifference are strictly imposed on the models~\cite{wen2021physics, klein2022polyconvex, tac2022data}.
In the recent work of Linden et. al.~\cite{linden2023neural}, they have proposed the use of a physics-augmented neural network (PANN), which can ensure that, besides material frame-indifference, physical constraints such as different material symmetry properties and volumetric growth conditions, can also be satisfied by construction. 
In the meantime, PANN makes use of the input convex neural network~(ICNN)~\cite{amos2017input} to ensure the polyconvexity of the energy storage function, a condition to guarantee the governing PDEs of finite elasticity are mathematically well-posed.
These examples of successfully using NN-based constitutive models for material modeling have shown that NNs have indeed great potential for characterizing materials with even more complex behavior. 
However, one important issue is that the training of NNs often require a large number of high-quality, diverse data points (i.e., diverse stress-strain states)~\cite{liu2021review, fernandez2021anisotropic, fuhg2022learning}.
Although such task can be easily done via numerical simulation for synthetic data, acquiring such data in using traditional experiments can be prohibitively expensive in terms of the cost of time, labor and resources.
This significantly hinders the application of NN-based constitutive models for characterizing real-world materials.

% fulfills all common physical constraints for hyperelastic materials, including objectivity, material symmetry, polyconvexity, growth conditions and thermodynamic consistency.
% Despite being able to address the issue of limited representational capability, a common drawback of the NN-based constitutive model is the increased dependency on data.
% The training of NNs requires the energy/stress-strain data distributed throughout the deformation space\cite{liu2021review, fernandez2021anisotropic, fuhg2022learning}.
% Acquiring such data through simple loading tests can be prohibitively expensive or nearly impossible. 
% This significantly hinders the application of NNs in identifying the constitutive models of real-world materials.

To address such difficulty, one possible way is to make use of the non-uniform full-field deformation data, which can be measured using techniques such as digital image correlation (DIC) or digital volume correlation (DVC)~\cite{marwala2010finite, pierron2012virtual, boddapati2023single}.
The recently proposed efficient unsupervised constitutive law identification and discovery~(EUCLID) framework~\cite{flaschel2021unsupervised, flaschel2022discovering, joshi2022bayesian} has shown to be highly effective by combining the full-field deformation data and the global reaction forces, which are also realistically measurable through ways of sparse regression.
%To circumvent the difficulty of obtaining stress data, some methods are dedicated to construct constitutive models using the non-uniform full-field displacement data, which contains a wide range of deformation states in the stress-strain space\cite{tung2024anti} and can be measured using digital image correlation (DIC) or digital volume correlation (DVC) technologies through a single test.
To leverage such data for training NN-based constitutive models, one approach is to set the loss function as the residuals of governing PDEs and the neural networks are trained in an unsupervised manner~\cite{huang2020learning, chen2021learning, thakolkaran2022nn, han2024learning}.
It is also possible to define the loss function as the difference between the model prediction and the measured data, and the training of the model can be formulated as a PDE-constrained optimization problem, termed as the hybrid FEM-NN approach~\cite{mitusch2021hybrid, zhang2022learning, tung2024anti}.
In this approach, the model prediction is obtained by numerically solving the governing PDEs under the same setting as the experiments and the gradients needed for the training process can be obtained via adjoint methods.
As compared to the former approach, although such approach can lead to larger computational cost and stronger requirement on the convergence of the numerical solver, as it has treated the governing PDEs as strict constrains for optimizing the model parameters, it can help increase the model accuracy, the optimization efficiency and possibly reduce the size of the neural networks~\cite{mitusch2021hybrid}.

In this work, we combine the hybrid FEM-NN framework with the PANN architecture~\cite{linden2023neural} for efficient construction of the constitutive models for isotropic hyperelastic materials.
By using a PANN-based architecture as our base model, we not only ensure that the necessary physical constraints are satisfied, but more importantly ensure the convergence of the numerical solver, which is crucial for the hybrid FEM-NN framework.
Following the hybrid FEM-NN framework, we then train the model using synthetic data obtained from numerical simulation by solving the corresponding PDE-constrained optimization problem.
The performance of the proposed method is extensively tested against various models, and in all cases, the three-dimensional isotropic constitutive models can be accurately and efficiently constructed from the two-dimensional full-field displacement and the total external load data obtained from a sample whose deformation is spatially non-uniform.
Such data can be easily measured by unaxial tension test for samples with geometric inhomogeneities.
We also test the predictability of the model for unseen geometries and loading conditions to verify the extrapolation ability of the optimized model.
Since we have treated the governing PDEs as strict constrains during the training process, the optimized model are solvable using FEM, and thus making it convenient to be implemented in commercial or open-source FEM packages.
%A small improvement was made on the PANN framework to enable the ICNNs to automatically satisfy the stress-free condition, i.e., yielding zero stress response in the undeformed state.
%As a physically consistent model, the PANN satisfies certain physical laws that ensure the NNs can be smoothly trained within the hybird FEM-NN framework.
%Through numerical simulation, we extensively test the performance of the hybrid FEM-NN framework using various analytical constitutive models.
%In all test cases, the three-dimensional isotropic constitutive models can be accurately and efficiently identified from the two-dimensional full-field displacement and external force data generated from a single uniaxial tension test.
%Examined under some simple deformation states, the learned constitutive models demonstrate good extrapolation capability beyond the training deformation modes.

The rest of the paper is organized as follows.
We start by introducing the basic formulation of our method, that is to formulate the construction of optimal constitutive models as a PDE-constrained optimization problem.
In the next section, we provide the details of the neural network architectures and the training and validation process for finding the optimal model. 
The results are shown in Section 4, and we conclude in Section 5.
%In Section 2, we present the mathematical formulation of the PDE-constrained optimization problem. 
%Following, we introduce the modified PANN architecture and the training method used in the FEM-NN framework in Section 3.
%In Section 4, we use the hybrid FEM-NN method to identify five classical analytical isotropic hyperelastic constitutive models and extensively evaluate the performance of the trained models beyond the training scenarios.
%Finally, a conclusion is made in Section 5.
Standard notation is used throughout the paper.
Normal fonts are used for scalars, and boldface fonts for vectors and second-order tensors.
Tensor and vector components are written with respect to a fixed Cartesian coordinate system with orthonormal basis $\lbrace \ee_i\rbrace_{i=1}^{3}$.
The summation convention is used for repeated Latin indices, unless otherwise indicated.
We denote by $\Ib$ the second-order identity tensor.
The prefixes tr and det indicate the trace and the determinant and superscript ${\top}$ the transpose of a second-order tensor.
Let ($\boldsymbol a$, $\boldsymbol b$) be vectors, ($\boldsymbol A$, $\boldsymbol B$) be second-order tensors and $\nabla$ the gradient operator.
We define the following: $\boldsymbol a \cdot \boldsymbol b = a_{i} b_{i}$, $(\boldsymbol A \cdot \boldsymbol a)_{i} = A_{il} a_{l}$, $(\boldsymbol A \cdot \boldsymbol B)_{il} = A_{ip}B_{pl}$, 
$\boldsymbol A: \boldsymbol B = A_{il}B_{il}$, $(\nabla \boldsymbol a)_{il} = \partial_l a_i$, $\nabla \cdot \boldsymbol a = \partial_i a_i$ and $ (\nabla \cdot \boldsymbol A )_{i} = \partial_{l} A_{il}$.
We denote by $H^k(\Omega)$ the Sobolev space $W^{k,2}(\Omega)$.
Note that boldface font is also used for matrix/vector assembled by the finite element method, e.g., a vector of nodal values $\Vb$ or a matrix~$\Ab$.
We use $\Ab^*$ to denote the adjoint of $\Ab$.
They are also used to denote vector/matrix used in the construction of neural networks, e.g., an input vector $\x$ or a weight matrix $\Wb$.
Our code can be found in \href{https://github.com/link577/NNhyper}{https://github.com/link577/NNhyper}.

\section{Problem formulation}

\subsection{Finite deformation elasticity}
\label{Sec:workflow}

Consider an elastic body that occupies a region ~${\mathbb{B}_0 \subset \mathbb{R}^3}$, at its stress-free reference state.
We denote by $\partial \mathbb{B}_0$ the boundary of the body and $\Nb$ the outward normal of $\partial \mathbb{B}$.
The boundary $\partial \mathbb{B}_0$ is assumed to be composed of two parts $\partial \mathbb{B}_0^D$ and $\partial \mathbb{B}_0^N$ such that~${\partial\mathbb{B}_0^N \cup \partial\mathbb{B}_0^D = \partial \mathbb{B}_0}$ and~${\partial\mathbb{B}_0^N \cap \partial\mathbb{B}_0^D = \emptyset}$.
Suppose that at time $t$, in response to some displacement $\bar{\ub}$ on $\partial \mathbb{B}_0^D$, and traction loading $\tb$ on $ \partial \mathbb{B}_0^N$, the body deforms and occupies a spatial region~${\mathbb{B}_t \subset \mathbb{R}^3}$.
The deformation map~${\boldsymbol{\varphi}_t: \mathbb{B} \rightarrow \mathbb{B}_t}$ sends a material point~${\X \in \mathbb{B}}$ to its spatial counterpart~${\x \in \mathbb{B}_t}$, i.e.,~${\x = \boldsymbol{\varphi}(\X, t)}$. 
The corresponding displacement field is defined as~${\ub(\X, t) = \x - \X}$.
For a hyperelastic material (in the absence of body force), consider the following total potential energy :
\begin{align} \label{Eq:lagrangian}
   \Pi[\ub]  =  \int_{\mathbb{B}} W(\Fb) \textrm{ d}\X -  \int_{\partial\mathbb{B}_0^{N}} \tb \cdot \ub \textrm{ d} S,
\end{align}
where $\Fb = \frac{\partial \x}{\partial \X} = \Ib + \frac{\partial \ub}{\partial \X}   $ is the deformation gradient, and $W$ is the strain energy density function (per volume in the reference configuration).
The governing partial differential equation (PDE) corresponding to the stationary point of Eq.~(\ref{Eq:lagrangian}) is:
\begin{equation} \label{Eq:BVP-strong}
\textrm{Div }  \Pb  = 0 \quad \textrm{in}  \, \, \mathbb{B}_0,
\end{equation}
with the following boundary conditions:
\begin{align}
    \label{Eq:BVP-ubc}
    \ub = \bar{\ub} &  \quad\textrm{on}  \, \, \partial\mathbb{B}_0^D  \\
     \label{Eq:BVP-tbc}
    \Pb \cdot \Nb = \tb  & \quad \textrm{on} \, \, \partial\mathbb{B}_0^N,
\end{align}
where the divergence operator is defined in the reference configuration $\mathbb{B}_0$, and the first Piola-Kirchhoff stress $\Pb$ is given by~${\Pb = \frac{\partial W}{\partial \Fb}}$ due to thermodynamic consistency.

An approximation of such stationary point can be found by first building a finite element mesh over the whole material domain and numerically solving the following weak form: find $\ub\in\mathcal{U}_h$ such that $\forall \vb\in\mathcal{V}_h$,
%To get the weak form, we take the variation of $\mathcal{E}$ with respect to $\ub$ in the direction of $\vb$.
\begin{align} \label{Eq:weak_form_fem}
    \delta \Pi[\ub; \vb] =  \int_{\mathbb{B}_0} \Pb(\Fb) : \nabla_{0}\vb   \textrm{ d}\X - \int_{\partial\mathbb{B}_0^{N}}\tb\cdot\vb\textrm{ d} S = 0,
\end{align} 
where we know: 
\begin{subequations}
\begin{align}
     \mathcal{U}_h \subset \mathcal{U} &= \big\lbrace \ub \in H^1(\Omega_0) \,\, \big| \,\, \ub = \bar{\ub} \textrm{ on } \partial\mathbb{B}_0^{D} \big\rbrace,  \\
    \mathcal{V}_h \subset \mathcal{V} &= \big\lbrace \vb \in H^1(\Omega_0) \,\, \big| \,\, \vb = \boldsymbol{0}\textrm{ on } \partial\mathbb{B}_0^{D} \big\rbrace.
\end{align}
\end{subequations}
The  approximate solution can be found via standard finite element method (FEM)~\cite{hughes2012finite}. 

The key component that governs the material behavior in finite deformation elasticity is the strain energy density function $W(\Fb)$, which needs to be inferred and determined from experimental measurements. In the following subsection, we will provide a detailed procedures of systematically obtaining such relationship by solving a PDE-constrained optimization problem.

\subsection{The PDE-constrained optimization framework}

Suppose that the strain energy density $W$, which we assumed to parameterized by some $\thetab$, i.e., $W = W(\Fb, \thetab)$. Such parameters can be the initial shear modulus as in the neo-Hookean model\cite{rivlin1948large}, or weights and biases for the neural-network-based constitutive models.
For a given set of parameters $\thetab$, one can in principle predict the material response to some loading conditions by solving Eq.(\ref{Eq:BVP-strong}). 
Ideally, one can also conduct a experiment under such loading, and collect the measured material response as dataset $\mathcal{D}$. 
Therefore, if we denote by $J$ a loss function to measure the difference of prediction and measurement, then the characterization of material behavior can be formulated as the following PDE-constrained optimization problem:
\begin{equation}
\begin{split}
   &  \text{find } \thetab, \   \text{to minimize} \   J[\ub(\thetab), \thetab], \ \text{given} \ \mathcal{D} \\
   & \text{with $\ub$ the solution of}\  \text{Eq}.~(\ref{Eq:BVP-strong}-\ref{Eq:BVP-tbc} ).  
    \end{split}
\label{eq:PDE-optimization}
\end{equation}

Note that conventionally, the experiments are often carried out using a geometry where rather uniform deformation or stress state can be achieved, such as uniaxial tension, biaxial tension, etc.
By assuming some known form of $W$, one can work out the analytic form of the force-displacement relation. Using data fitting method, one can then determine the optimal parameters.
However, such procedures highly rely on the pre-assumed form of the constitutive models. In addition, in order to assure that the constitutive models inferred from the experiments are valid in complex stress-strain states, ideally, one will need to conduct measurements under many different stress-strain pairs. This is not only costly in time, economy and human resources, but more importantly, some experiments can be difficult to be carried out. 

In order to address this challenge, we propose to conduct experimental measurement on a specimen where the deformation field is non-uniform, e.g., a thin membrane with complex geometry, a block under indentation, etc.
Under such setting, using one single experiment, a considerable number of different stress-strain states can be explored on different material points. 
In the meantime, in such setting working out the analytic form of force-displacement relation for the prediction of material response can be extremely difficult, but we can turn to numerical methods such as FEM to serve for the purpose. 
Let $\Ub=\lbrace {u}_l \rbrace \in\mathbb{R}^L$ be the vector of solution nodal displacement values with $L$ the degrees of freedom of the finite element problem.
We denote the corresponding algebraic equation of Eq.~(\ref{Eq:weak_form_fem}) by $\Gb(\Ub, \thetab) = 0$.
Then, the core of our method is to:
\begin{equation}
\begin{split}
    & \text{find } \thetab, \    \text{to minimize} \   J[{\boldsymbol{U}}(\boldsymbol{\theta}), \boldsymbol{\theta}]\  \text{given} \  \mathcal{D} \\
    & \text{with $\Ub$ the solution of}\    \boldsymbol{G}({\boldsymbol{U}}, \boldsymbol{\theta}) = \boldsymbol{0},
\end{split}
\label{eq:PDE-constrained_optimization}
\end{equation}

To capture the diverse stress-strain states, we can use the full-field displacement data ${\hat{\ub}}$ as part of our data, which can be measured using techniques such as digital imaging correlation (DIC). 
However, stress is rather difficult to be measured point by point, even on the boundary, and therefore, we use the total external load $\hat{\fb}$, which can be measured in a much more convenient fashion as another source of our data.
As a result, $\mathcal{D} = \lbrace \hat{\ub}_j, \hat{\fb}_j\rbrace_{j=1}^{N}$.
Let us consider the following loss function:
\begin{equation}
    J = \sum_{j=1}^{N} \left(\frac{1}{2V}\int_{V} \left\lVert \frac{\tilde{{\boldsymbol{u}}}_j - \hat{\ub}_j}{u_{\text{max}}} \right\rVert^2 {\rm d} V + \frac{1}{2} \left\lVert \frac{\tilde{{\fb}}_j -\hat{\fb}_j}{f_{\text{max}}}\right\rVert^2\right)
\label{eq:loss_function},
\end{equation}
where $u_\text{max}$ and $f_\text{max}$ are the maximum displacement and force components from the measured data, $\tilde{\ub}$ and $\tilde{\fb}$ are the displacement field and external force determined by the solution nodal values $\Ub$, and $V$ the region of interest (ROI) over which the displacement field is measured.
The optimization problem (\ref{eq:PDE-constrained_optimization}) can be solved via the adjoint method, which will be discussed in the next section.

\section{Methodology}

\subsection{Adjoint optimization}

\begin{figure}[!h]
\centering
\includegraphics[width=\textwidth]{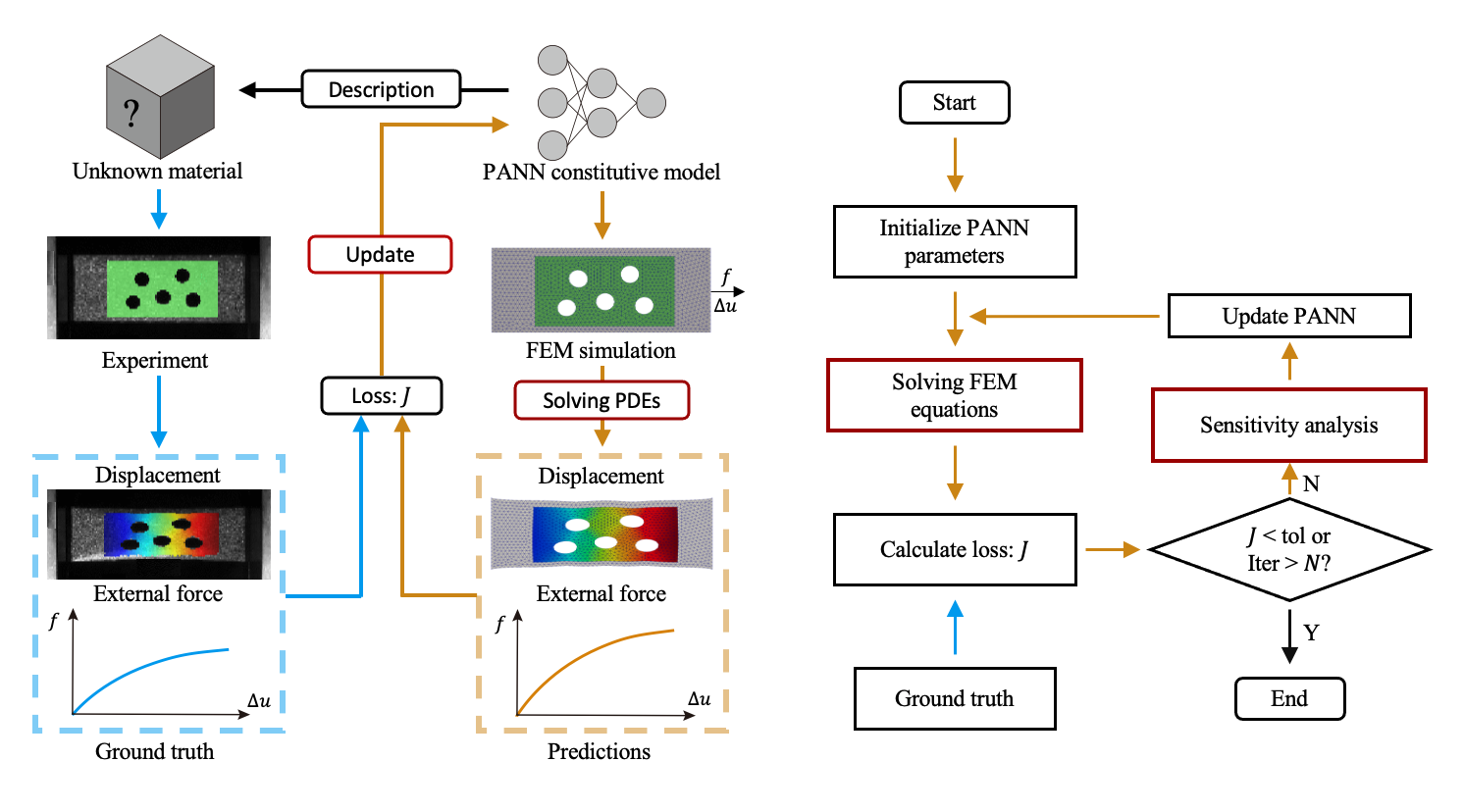}
\caption{Schematic of the hybrid finite element method-neural network~(FEM-NN) optimization framework used to solve the PDE-constrained optimization problems.}
\label{fig:framework}
\end{figure}

The core of solving the optimization problem (\ref{eq:PDE-constrained_optimization}) is to calculate the gradient of the loss function with respect to the parameters $\thetab$ under the constrains of the governing PDEs. 
Once this gradient can be efficiently calculated, we can follow the procedures shown in in Fig.~\ref{fig:framework} to iteratively obtain the optimal $\thetab$.
We first initialize the system with some parameter $\thetab_0$, which is supplied to the FEM module to calculate the predicted displacement field $\tilde{\ub}(\thetab_0)$ by solving $\Gb(\Ub, \thetab_0) = 0$.
Then, such predicted field is used to evaluate the loss function $J[{\tilde{\boldsymbol{u}}}(\boldsymbol{\theta}_0), \boldsymbol{\theta}_0]$, followed by computing its gradient with respect to the parameters $\thetab$, i.e., $\frac{{\rm d}J}{{\rm d}\boldsymbol{\theta}} \big|_{\thetab=\thetab_0}$.
Following such gradient, we can then update the parameters $\thetab_0$ using the BFGS optimization method~\cite{broyden1970convergence} to minimize the value of $J$.
Above operations are repeated iteratively until the BFGS algorithm converges or the number of iterations exceeds a threshold.

In this work, we make use of the adjoint method to compute the gradient needed for efficient optimization algorithms. 
We will perform discretization first, and then apply the adjoint method to the discrete set of equations.
Within the discrete finite element function space, the gradient of loss function $J$ can be determined by the following relation:
\begin{equation}
    \frac{{\rm d}J}{{\rm d}\boldsymbol{\theta}} = \frac{\partial J}{\partial \Ub}\frac{{\rm d}\Ub}{{\rm d}\boldsymbol{\theta}} + \frac{\partial J}{\partial \boldsymbol{\theta}},
\label{eq:sensitivity_analysis}
\end{equation}
where the calculation of  $\frac{\partial J}{\partial \Ub}$ and $\frac{\partial J}{\partial \boldsymbol{\theta}}$ is straightforward.
On the other hand, the parameters $\thetab$ and displacement field $\ub$ are implicitly related via equilibrium equation, i.e. $\Gb(\Ub, \thetab) = 0$.
By differentiating both sides, we have the following:
\begin{equation}
    \frac{\partial\boldsymbol{G}}{\partial \Ub} \frac{{\rm d}\Ub}{{\rm d}\boldsymbol{\theta}} = -\frac{\partial \boldsymbol{G}}{\partial \boldsymbol{\theta}},
\label{eq:tangent_linear_equation}
\end{equation}
therefore by chain rule, we have:
\begin{align} 
\label{Eq:chain_rule}
    \frac{\textrm{d}{J}}{\textrm{d}\thetab} =  -{\underbrace{ {\frac{\partial J}{\partial \Ub} } \Big(\frac{\partial\Gb}{\partial\Ub}\Big)}_{\lambda^*}}^{-1} \frac{\partial \Gb}{\partial \thetab} + 
    \frac{\partial J}{\partial \thetab}. 
    \end{align}

In principle, there are two ways to calculate $\frac{{\rm d}J}{{\rm d}\boldsymbol{\theta}}$.
The first is to solve the tangent linear equations in Eq.~\ref{eq:tangent_linear_equation}, and substitute the solutions $\frac{{\rm d}\Ub}{{\rm d}\boldsymbol{\theta}}$ into Eq.~\ref{eq:sensitivity_analysis}.
The second is to solve the following adjoint equation:
\begin{equation}
    \frac{\partial\boldsymbol{G}}{\partial \boldsymbol{{\boldsymbol{U}}}}^{*}  \boldsymbol{\lambda} = \frac{\partial J}{\partial {\boldsymbol{U}}}^* \\
\label{eq:adjoint_equation},
\end{equation}
and substitute the adjoint variables $\lambdab$ to Eq.~\ref{Eq:chain_rule}, and such approach is often called the adjoint method.
Since the dimension of the parameters $\boldsymbol{\theta}$ is much larger than the dimension of the loss function $J$~(a scalar in this work), the adjoint method is more efficient than the first method.
Open source library \textit{dolfin-adjoint}~\cite{mitusch2021hybrid, mitusch2019dolfin} is used in this work to solve the adjoint equations.

\subsection{Physics-augmented neural network architecture}

In this work, we adopt a data-driven approach for the extraction of the optimal constitutive model. 
But before we start any optimization process, we need to first ensure that the constitutive model is physically sensible by construction.
As addressed in \cite{linden2023neural}, several important restrictions must be satisfied.
 
First, the principle of objectivity~\cite{truesdell2004non} states that the mechanical response of material is independent of the observer. For hyperelastic materials, such restrictions can be written as:
\begin{equation}
{W}(\boldsymbol{F}) = {W}(\boldsymbol{Q} \boldsymbol{F}), \ \forall \boldsymbol{Q} \in \text{SO(3)}
\end{equation}
with $\text{SO(3)}$ the special orthogonal group. 
%{\bf need to add a few words here}

Second, material symmetry requires $W$ to be invariant under the symmetry transformation within the corresponding material symmetry group $\mathcal{G} $\cite{haupt2002continuum}, i.e., 
\begin{equation}
{W}(\boldsymbol{F}) = {W}(\boldsymbol{F} \boldsymbol{Q}), \  \forall \boldsymbol{Q} \in \mathcal{G}.
\label{eq:material_symmetry}
\end{equation}
Specifically, for the isotropic material, $\mathcal{G} = \text{SO(3)}$.
%{\bf need to add a few words here}

In addition, in this work we consider a material without pre-stress, and therefore,  
\begin{equation}
{\boldsymbol{P}}(\Ib) = \frac{\partial{W}}{\partial\boldsymbol{F}}\Big|_{\boldsymbol{F}=\Ib} = \boldsymbol{0}.
\end{equation}
In the meantime, to ensure balance of angular momentum, we also need to ensure that:
\begin{equation}
\Pb \Fb^T = \Fb \Pb^T.
\end{equation}

Moreover, in order to ensure that the solution of of Eq.~\ref{Eq:weak_form_fem} exist, we follow the work of~\cite{ball1976convexity} and impose a polyconvex condition for $W$. 
The energy storage function ${W}(\boldsymbol{F})$ is polyconvex if and only if there exists a function $\mathcal{P}: \mathbb{R}^{3\times 3}\times \mathbb{R}^{3\times 3}\times  \mathbb{R}^3 \to \mathbb{R} $ such that:
\begin{equation}
{W}(\boldsymbol{F}) = \mathcal{P}(\boldsymbol{F}, \text{cof}\boldsymbol{F},  \text{det}\boldsymbol{F}) ,\  \  \   \mathcal{P} \text{ is convex in its arguments.}
\end{equation}

Besides, volumetric growth condition, i.e., ${W}(\boldsymbol{F}) \to \infty,\ \text{as} \ \text{det}\boldsymbol{F} \to 0^+$ is also considered to reflect the fact that an infinite energy is required for an infinitely large volumetric compression.
    
In order to meet all the aforementioned constraints, we follow the work of \cite{linden2023neural} and adopt a modified physical-augmented neural network~(PANN) architecture in this work.

\begin{figure}[!h]
\centering
\includegraphics[width=\textwidth]{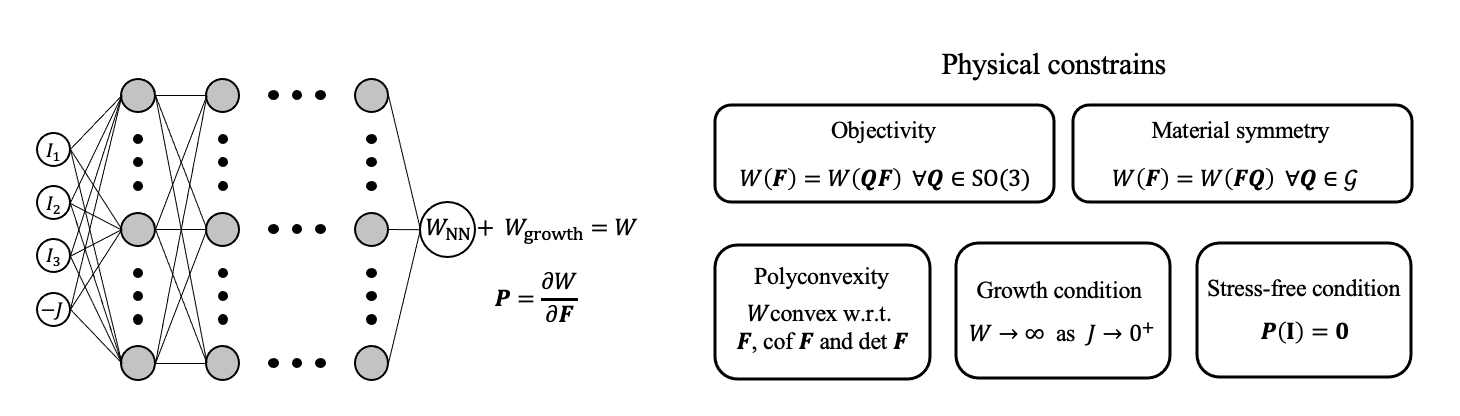}
\caption{The architecture of the modified PANN.}
\label{fig:network}
\end{figure}

As shown in Fig.~\ref{fig:network}, we decompose the total strain energy density $W$ into two parts: the neural network part $W_{\text{NN}}$ and an additional term $W_{\text{growth}}$.
The additional term $W_{\text{growth}}$ is introduced to guarantee the volumetric growth condition:
\begin{equation}
    {W}_{\text{growth}} = \theta_{\text{g}}(J-\text{ln}J),\;\;\theta_{\text{g}}\geq 0
\label{eq:manual_term}.
\end{equation}
Here, $\theta_{\text{g}}$ is an additional non-negative scalar parameter besides the weights and bias in ${W}_{\text{NN}}$, which also needs to be optimized.

As for ${W}_{\text{NN}}$, the input of this network is chosen as: $\boldsymbol{x} = [I_1, I_2, I_3, -J]^T$, where $I_i$ represents the $i$th principal invariant of the right Cauchy-Green tensor $\boldsymbol{C} = \boldsymbol{F}^T  \boldsymbol{F}$, and $J = \text{det}\boldsymbol{F}$ is the determinant of the deformation gradient. 
Using such invariant-based inputs, we can guarantee that the NN-based constitutive model obeys objectivity, balance of angular momentum as well as isotropic material symmetry automatically~\cite{linden2023neural}.

In addition, we adopt an input convex neural network~(ICNN)~\cite{amos2017input}, where the output of the network is a convex function of its inputs, as our architecture for the neural networks. 
In such a network, the activation functions need to be node-wise convex and non-decreasing, and all the weights, not including the biases, need to be non-negative. 
In such a way, we can guarantee the poly-convexity condition.
In this work, we have used \emph{softplus} as our activation function.

% Note that, as pointed out by\cite{linden2023neural}, the activation function that acts on $J$ must be convex, but not necessarily non-decreasing, and therefore, we include the additional term $-J$ here in the input, which is important for the NN-based constitutive models\cite{klein2022polyconvex}.
% Meanwhile, such choice is also useful to impose the stress-free conditions.

Note that, we include an additional term $-J$ in the input, which is useful to impose the stress-free conditions that will be introduced next.
Let us denote the output of the input layer of the ICNN as $\boldsymbol{y}^1$, i.e. $\boldsymbol{y}^1 = \boldsymbol{W}^1\boldsymbol{x} + \boldsymbol{b}^1$, where $\boldsymbol{W}^1 \in \mathbb{R}^{n_1\times 4}$ and $\boldsymbol{b}^1 \in \mathbb{R}^{n_1}$ are the weights and bias of the first layer respectively. $n_1$ denotes the number of neurons in the first hidden layer.
Since $W_{\rm growth}$ automatically satisfies the stress-free condition at $\Fb = \Ib$, the stress free condition only needs to be imposed on $W_{\rm NN}$:
\begin{equation}
    \frac{\partial{W}_{\text{NN}}}{\partial\boldsymbol{F}}\Big|_{\boldsymbol{F}={\Ib}} = \frac{\partial{W}_{\text{NN}}}{\partial\boldsymbol{y}^1}\frac{\partial\boldsymbol{y}^1}{\partial\boldsymbol{x}}\frac{\partial\boldsymbol{x}}{\partial\boldsymbol{F}}\Big|_{\boldsymbol{F}=\Ib} = \boldsymbol{0}
\label{eq:stress_free_consition_NN}.
\end{equation}
Such condition is guaranteed if the following relation is satisfied:
\begin{equation}
    \frac{\partial\boldsymbol{y}^1}{\partial\boldsymbol{x}}\frac{\partial\boldsymbol{x}}{\partial\boldsymbol{F}}\Big|_{\boldsymbol{F}=\Ib} = 
    \boldsymbol{W}^1\frac{\partial\boldsymbol{x}}{\partial\boldsymbol{F}}\Big|_{\boldsymbol{F}=\Ib} = \boldsymbol{0}
\label{eq:sufficient_condition_stree_free},
\end{equation}
For the specific input vector $\boldsymbol{x}$ used in this work, we have:
\begin{equation}
    \frac{\partial I_1}{\partial\boldsymbol{F}}\Big|_{\boldsymbol{F} = \Ib} = 2\Ib,\;\;
    \frac{\partial I_2}{\partial\boldsymbol{F}}\Big|_{\boldsymbol{F} = \Ib} = 4\Ib,\;\;
    \frac{\partial I_3}{\partial\boldsymbol{F}}\Big|_{\boldsymbol{F} = \Ib} = 2\Ib,\;\;
    \frac{\partial (-J)}{\partial\boldsymbol{F}}\Big|_{\boldsymbol{F} = \Ib} = -\Ib
\label{eq:derivative_of_invariant}.
\end{equation}
Now, let $\boldsymbol{w}^{i} \in \mathbb{R}^{n_1}$ represents the $i_\textrm{th}$ column of $\boldsymbol{W}^1$, i.e., $\Wb^1 = [\wb^1, \wb^2, \wb^3, \wb^4]$.
Then, by simply imposing: 
\begin{equation}
     \boldsymbol{w}^{4} = 2\boldsymbol{w}^{1} + 4\boldsymbol{w}^{2} + 2\boldsymbol{w}^{3}
\label{eq:constrain_column_vector_stress_free},
\end{equation}
we can ensure that Eq.~(\ref{eq:sufficient_condition_stree_free}) is satisfied.
% So, Eq.~\ref{eq:sufficient_condition_stree_free} is equal to $\boldsymbol{W}^1\boldsymbol{v} = \boldsymbol{0}$, with $\boldsymbol{v} = [2,4,2,-1]^T$, which can be satisfied by constrain the column vectors of $\boldsymbol{W}^1$ as:

% where, 
% With another part $W_\text{growth}$~(Eq.~\ref{eq:manual_term}) is set to be stress-free, the modified PANN architecture satisfies the stress-free conditions by construction.

% Note that the partial derivatives of the invariant-based inputs w.r.t $\boldsymbol{F}$ for undeformed state are linearly dependent, i.e.,
% \begin{equation}
%     \frac{\partial I_1}{\partial\boldsymbol{F}}\Big|_{\boldsymbol{F} = \Ib} = 2\Ib,\;\;
%     \frac{\partial I_2}{\partial\boldsymbol{F}}\Big|_{\boldsymbol{F} = \Ib} = 4\Ib,\;\;
%     \frac{\partial I_3}{\partial\boldsymbol{F}}\Big|_{\boldsymbol{F} = \Ib} = 2\Ib,\;\;
%     \frac{\partial (-J)}{\partial\boldsymbol{F}}\Big|_{\boldsymbol{F} = \Ib} = -\Ib
% \label{eq:derivative_of_invariant}.
% \end{equation}
% As a sufficient condition to satisfy the requirement of a stress-free condition, we set the column vectors of the weight matrix in the first layer of the ICNN as:
% \begin{equation}
%      \boldsymbol{w}_{4} = 2\boldsymbol{w}_{1} + 4\boldsymbol{w}_{2} + 2\boldsymbol{w}_{3}
% \label{eq:stress_free_sufficient_consition_specify},
% \end{equation}
% where, $\boldsymbol{w}_{i}$ represents the $i_\textrm{th}$ column of the weight matrix.

\section{Results} \label{sec:numerical_results}

In this section, we use the modified PANN architecture as our constitutive model, and follow the adjoint optimization procedures to obtain the optimal model for isotropic hyperelastic materials.
The training data is generated by conducting FEM simulation on a 2D thin plate with randomly placed holes using different hyperelastic models, which aims at simulating the experimental settings. 
Then, the optimization problem (\ref{eq:PDE-constrained_optimization}) is solved iteratively by following the procedures in Fig.~\ref{fig:framework}, followed by the validation and verification of the optimized model in unseen geometry, unseen loading conditions, etc.

\subsection{Data generation}

\begin{figure}[!h]
\centering
\includegraphics[width=0.5\textwidth]{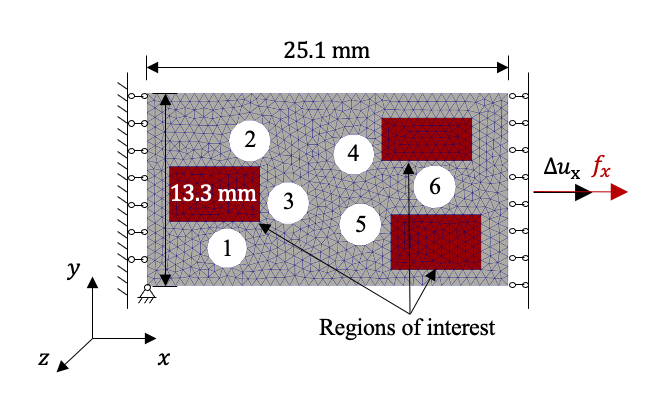}
\caption{
FEM setups of the 2D plates for generating the ground truth displacement field and external force data for training.
The red regions denote the regions of interest where displacement field is measured.
}\label{fig:data_generation_FEM}
\end{figure}

As mentioned earlier, to produce as diverse stress-strain pairs as possible in one experiment, a non-uniformly deformed sample is needed.
Therefore, we have chosen a 2D plate with randomly distributed circular holes as shown in Fig.~\ref{fig:data_generation_FEM}.
The length of the plate is chosen as 25.1 mm, and the width 13.3 mm. 
The radii of the six circular holes on the training sample, numbered 1 to 6, are 1.39~\text{mm},1.45~\text{mm}, 1.47~\text{mm}, 1.43~\text{mm}, 1.47~\text{mm and }1.49~\text{mm} respectively.
Plane stress assumption is adopted in the simulation.
% The thickness is regarded as much smaller than other two dimensions, and therefore, we adopt a plane stress assumption in the simulation. 

During the simulation, the left end of the sample was fixed in $x$-direction while the right end was stretched from $\Delta u_x = 0~\text{mm}$ to $5~\text{mm}$~(approximately 20\% axial strain) in $x$-direction through 21~($i = 0,1,\cdots, 20$) uniform loading steps.
The left-bottom corner was fixed to avoid possible rigid-body motion.
For each loading step, we calculate the displacement field as well as the total reaction force $f_x$ on the right boundary.
Note that in practical DIC measurements, boundary effects will lead to inaccuracies in displacement measurements at the edges of the sample~\cite{chu1985applications, sutton2009image}.
% Note that in reality, usually only several regions of interest (ROIs) will be picked for measuring the full-field displacement, rather than over the whole specimen. 
In light of this, only displacement data from three regions of interest (ROIs) are used for training our NN-based constitutive models, and these regions are highlighted in red in Fig.~\ref{fig:data_generation_FEM}. 
% we have chosen three rectangle regions in Fig.~\ref{fig:data_generation_FEM}, and they are highlighted using red color. 
% Only the displacement data in these red boxes will be used for the training of our NN-based constitutive models.
Among the data generated from the training sample, we use the data from nine loading steps, $j = \{1,2,3,10,11,12,18,19,20\}$ in Eq.~\ref{eq:loss_function}, for training, the others were used for validation.

Five classical isotropic hyperelastic models are chosen to generate training and validation data.
They are the neo-Hookean~(NH) model~\cite{rivlin1948large}:
\begin{equation}
\begin{split}
    W_{\text{NH}} &= \frac{\mu}{2}\left(I_1 - 3 - 2 \text{ln} J \right) + \frac{\lambda}{2}(J-1)^2,
\end{split}
\label{eq:NH}
\end{equation}
the Mooney-Rivilin~(MR) model~\cite{mooney1940theory}:
\begin{equation}
\begin{split}
    W_{\text{MR}} &= \frac{\mu}{2}(\bar{I}_2 - 3) + \frac{\kappa}{2}(J-1)^2,
\end{split}
\label{eq:MR}
\end{equation}
the five-term Arruda-Boyce~(AB) model~\cite{arruda1993three}:
\begin{equation}
\begin{split}
    W_{\text{AB}} &= \mu\left[\frac{1}{2}(\bar{I}_1 -3) + \frac{1}{20N}(\bar{I}_1^2 - 9) + \frac{11}{1050N^2}(\bar{I}_1^3 - 27) + \frac{19}{7000N^3}(\bar{I}_1^4 - 81) \right. \\
    &\left.+ \frac{519}{673750N^4}(\bar{I}_1^5 - 243) \right] + \frac{\kappa}{2}(J-1)^2,
\end{split}
\label{eq:AB}
\end{equation}
the Gent model~\cite{gent1996new}:
\begin{equation}
\begin{split}
    W_{\text{Gent}} &= -\frac{\mu J_m}{2}\, \ln\left(1 - \frac{\bar{I}_1-3}{J_m}\right) + \frac{\kappa}{2}\left(\frac{J^2-1}{2} - \text{ln}J\right),
\end{split}
\label{eq:Gent}
\end{equation}
and the Fung model~\cite{fung2013biomechanics}:
\begin{equation}
\begin{split}
    W_{\text{Fung}} &= \frac{\mu}{2b}(b(\bar{I}_1-3) + e^{b(\bar{I}_1-3)) -1} + \frac{\kappa}{2}(J-1)^2.
\end{split}
\label{eq:Fung}
\end{equation}
In Eq.~\ref{eq:NH}$\sim$\ref{eq:Fung} $\bar{I}_1 = J^{-2/3}I_1$, $\bar{I}_2 = J^{-4/3}I_2$ are the invariants of $\overline{\boldsymbol{C}} = J^{-2/3}\boldsymbol{C}$.
$\lambda$ and $\mu$ are the Lam$\acute{\text{e}}$ constants and $\kappa$ is the bulk modulus.
They are related to the Young's modulus $E$ and Poisson’s ratio $\nu$ by:  
\begin{equation}
\mu=\frac{E}{2(1+\nu)}, \quad \lambda=\frac{E\nu}{(1+\nu)(1-2\nu)}, \quad \kappa = \frac{E}{3(1-2\nu)}.
\end{equation}
The material parameters used for generating the data is summarized in Table~\ref{table:material_constants}.

\begin{table}[!h]
\centering
\begin{tabular}{cccccc}
\hline
 & $E$~(kPa)& $\nu$ & N &$J_m$ & b\\
\hline
NH & 100 & 0.3 & - & - & -\\
\hline
MR& 100 & 0.49 & - & - & -\\
\hline
AB & 100 & 0.3 & 10 & - & -\\
\hline
Gent& 100 & 0.3 & - & 0.2 & -\\
\hline
Fung & 100 & 0.3 & - & - & 1\\
\hline
\end{tabular}
\caption{Material parameters for the NH, MR, AB, Gent and Fung models}
\label{table:material_constants}
\end{table}

These five types of hyperelastic models are not only widely used ones, but can also capture distinct mechanical behaviors of hyperelastic materials. 
For example, Gent model is known to produce the strong strain-stiffening effect of polymers, which is controlled by the parameter $J_m$.
Here, we set the value for $J_m$ to 0.2 to induce strain-stiffening behavior at around 20\% strain, aimed at evaluating the effectiveness of our method.

%In this work, for Young's modulus $E=100\text{kPa}$ and Poisson's ratio $\nu = 0.3$,
%with $E=100\text{kPa}$, $\nu = 0.49$, were used to generate the synthetic data.
%Here, $J_m$ is set to 0.2.

%In the realm of material science, the NH, MR, and Gent models are pivotal in describing the nonlinear elastic behavior of rubber-like materials under various strain conditions.
%The NH model is favored for its simplicity and effectiveness in handling moderate deformations.
%The MR model is more complex and offering improved accuracy over a wider range of deformations.
%The Gent model excels at describing the behavior of materials under large strains by incorporating a limit stretch parameter $J_m$ which predicts stress saturation. This feature makes it particularly effective for materials that demonstrate significant strain hardening followed by softening.
%Selecting these three models for benchmarking facilitates a robust test of the {MINN}'s predictive capabilities, encompassing simple to complex, and moderate to extreme material responses, thereby providing a thorough assessment of its performance in predicting constitutive relations.

\subsection{Training, validation and test}

\begin{figure}[!h]
\centering
\includegraphics[width=0.45\textwidth]{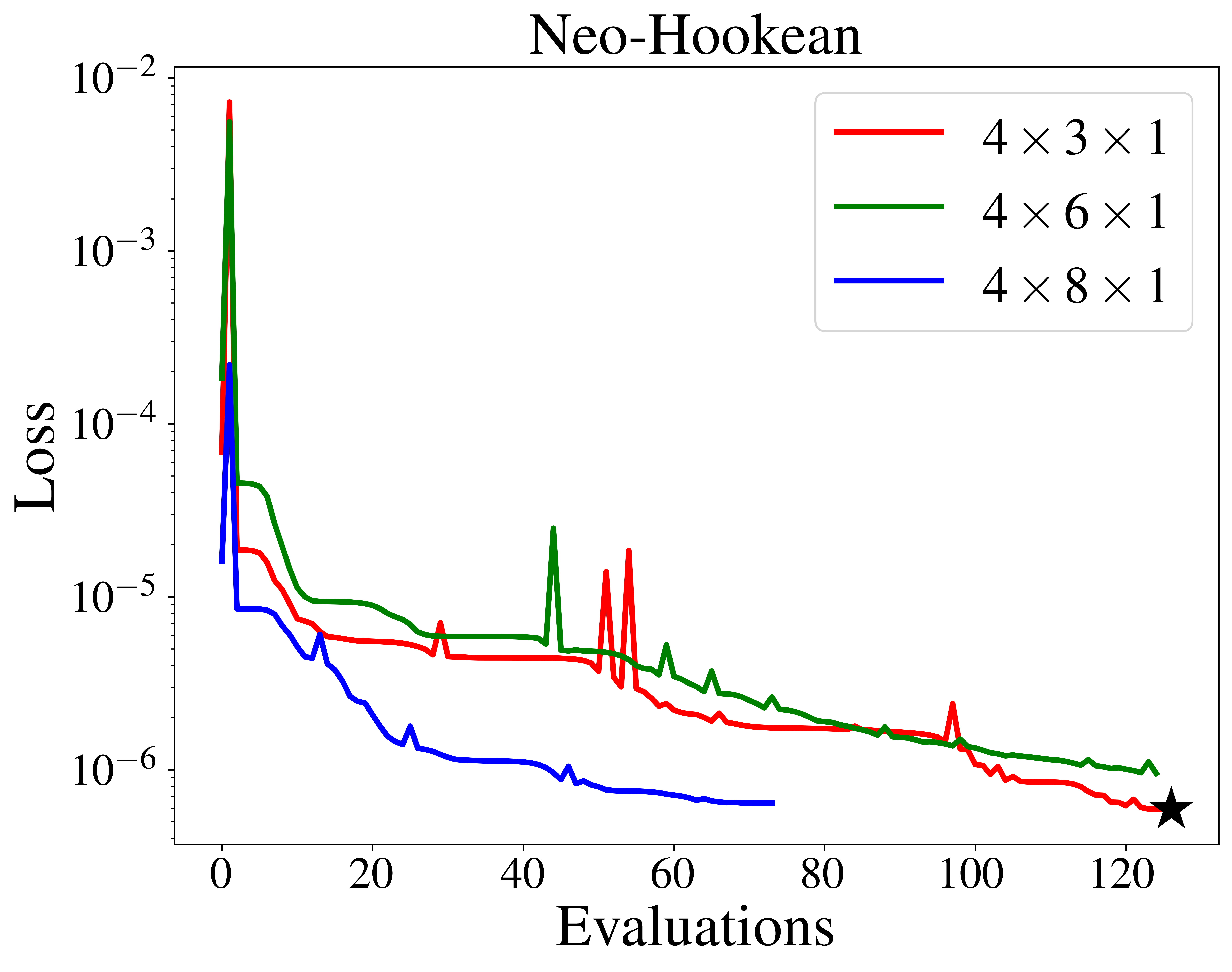}
\caption{
Training loss of three one-hidden-layer PANN models with 3, 6, and 8 neurons on the neo-Hookean model.
With the maximum number of iterations set to 100, all three models achieved loss values around $1\times 10^{-6}$, with the model containing 3 neurons having the lowest loss at $5.9\times 10^{-7}$.
During each iteration, multiple evaluations of the loss were performed until the loss was smaller than in the previous step.
}\label{fig:neo-Hookean_loss}
\end{figure}

A modified PANN architecture with one hidden layer is considered in this work, and three models, each with 3, 6 and 8 neurons in the hidden layer are studied.
Once the model is chosen, it is initialized with random parameters including weights and biases, and optimized following the procedures in Fig.~\ref{fig:framework}.
The training loss of the three NNs on NH model is shown in Fig.~\ref{fig:neo-Hookean_loss}.
All three models achieved comparable loss values around $1\times 10^{-6}$, with the model containing 3 neurons having the lowest loss.
Similar results can be observed in the other four constitutive models.
The results indicate that, due to the strict physical constraints imposed on the PANN itself and during the training process, a simple $4\times 3\times 1$ model is sufficient to accurately and efficiently (within 100 iterations) identify the considered constitutive models.
And therefore, in the following, we will present the results obtained from such model. 

\begin{figure}[!h]
\centering
\includegraphics[width=0.5\textwidth]{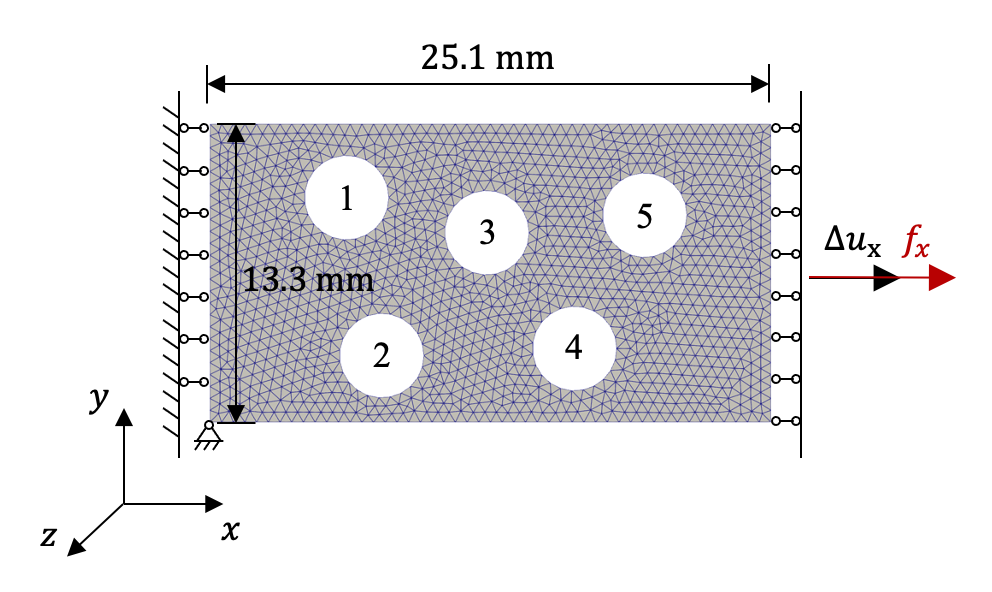}
\caption{
FEM setups of the 2D plates with unseen geometry for generating the ground truth displacement field and external force data for testing.
}\label{fig:test_sample}
\end{figure}

After training, the performance of the PANN constitutive models were first evaluated on the training sample and another unseen test sample by FEM.
The test sample is another $25.1~\text{mm}\times 13.3~\text{mm}$ 2D plate, with five randomly distributed holes, that are differently placed than the training sample.  
In the test sample, the radii of the five holes are all set to be 1.9~\text{mm}, as shown in Fig.~\ref{fig:test_sample}

\begin{figure}[!h]
\centering
\includegraphics[width=0.8\textwidth]{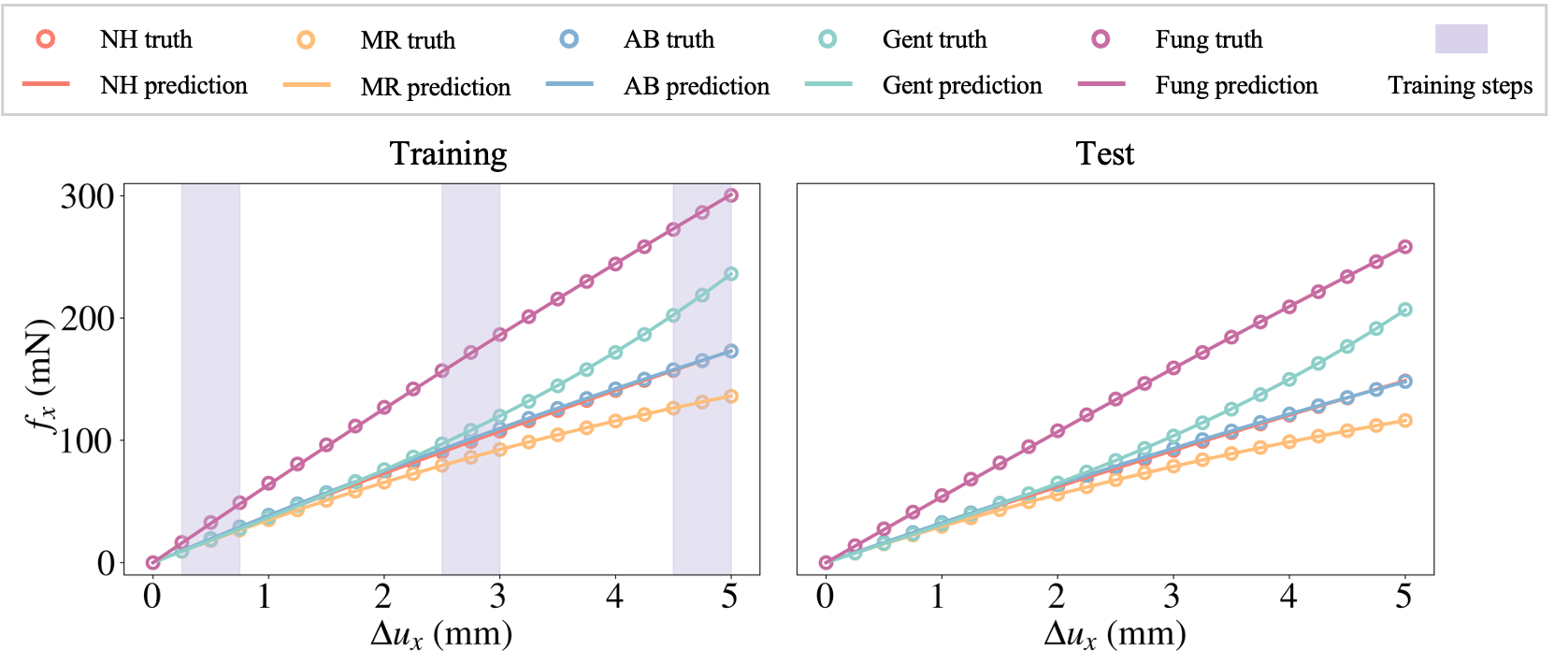}
\caption{
Comparisons of the force-displacement curves calculated by FEM with the ground truth and PANN constitutive models on the training~(left) and test~(right) samples.
}\label{fig:force_test}
\end{figure}

We first compare the force-displacement curves predicted from the optimized PANN models and the ground truth, which is shown in Fig.~\ref{fig:force_test}.
The circular dots are the ground truth data, where as the solid lines represent the prediction from the trained model. 
The light purple regions in the training results represent the nine loading steps used for training.
It can be seen that the force-displacement curves can be perfectly reproduced by the PANN models for all five different hyperelastic materials, even on the testing case, which is not seen by the model before. 
The maximum relative errors for force prediction is less than $2.9\%$.

\begin{figure}[!h]
\centering
\includegraphics[width=\textwidth]{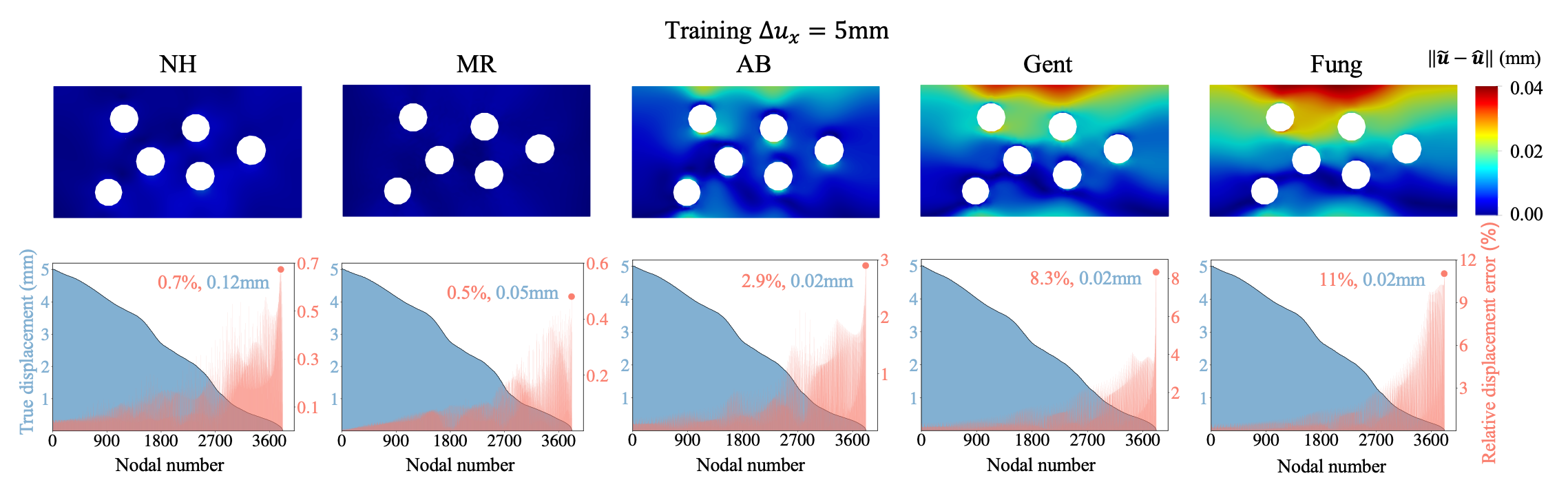}
\caption{Displacement error between the PANN prediction $\tilde{\boldsymbol{u}}$ and the ground truth $\hat{\ub}$ on the training sample with axial stretch $\Delta u_x = 5\text{mm}$.
The contour plots above shows the absolute error defined as $u_e = \lVert \tilde{\boldsymbol{u}}-\hat{\ub} \rVert$.
The figures below shows the relative displacement error on FEM nodes together with corresponding nodal displacement.
The relative error is calculated by $u_r = u_e/ (\lVert \hat{\ub} \rVert + u_0)$, where $u_0$ is set to $10^{-9}$mm to prevent the denominator from being zero.
}\label{fig:training_disp_error}
\end{figure}

\begin{figure}[!h]
\centering
\includegraphics[width=\textwidth]{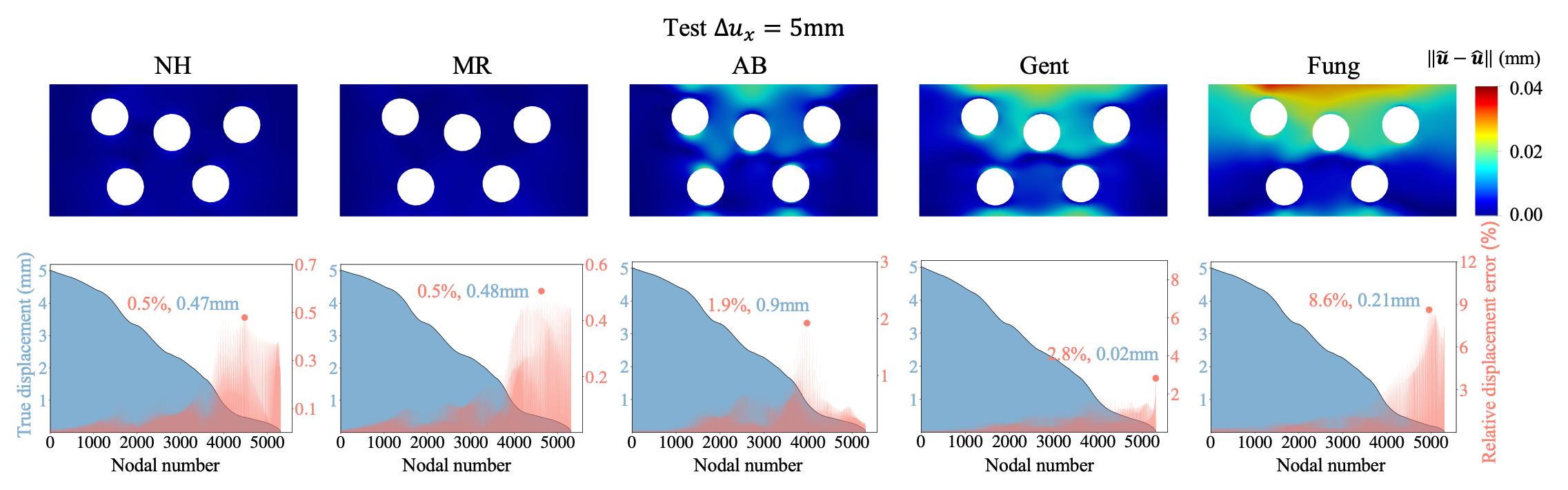}
\caption{Displacement error between the PANN prediction $\tilde{\boldsymbol{u}}$ and the ground truth $\hat{\ub}$ on the test sample with axial stretch $\Delta u_x = 5\text{mm}$.
The contour plots above shows the absolute error defined as $u_e = \lVert \tilde{\boldsymbol{u}}-\hat{\ub} \rVert$.
The figures below shows the relative displacement error on FEM nodes together with corresponding nodal displacement.
The relative error is calculated by $u_r = u_e/ (\lVert \hat{\ub} \rVert + u_0)$, where $u_0$ is set to $10^{-9}$mm to prevent the denominator from being zero.
}\label{fig:test_disp_error}
\end{figure}

We then compare the displacement field of the NN-based prediction $\tilde{\boldsymbol{u}}$ with the ground truth $\hat{\ub}$.
The contour plots of absolute displacement error ($u_e = \lVert \tilde{\boldsymbol{u}}-\hat{\ub} \rVert$) for the training and test samples at $\Delta u_x = 5\text{mm}$~($\sim 20\%$ axial stretch) are shown in Fig.~\ref{fig:training_disp_error} and Fig.~\ref{fig:test_disp_error} respectively.
As we can easily observe, the maximum displacement error occurred in the Fung model, approximately $0.04~\text{mm}$, which is much smaller than the overall stretch $\Delta u_x$.
The best results were obtained for the NH and MR models, with the maximum error being less than $0.01~\text{mm}$.

\begin{figure}[!h]
\centering
\includegraphics[width=0.45\textwidth]{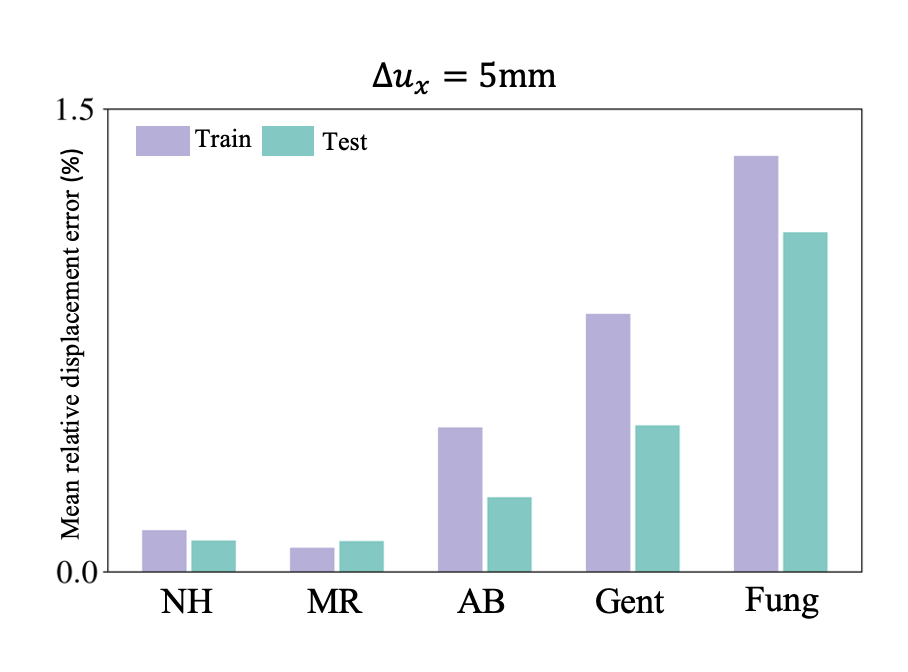}
\caption{
The full-field average relative displacement error on the training and test samples at an axial stretch $\Delta u_x = 5\text{mm}$.
%The error is calculated by $\bar{u}_r = \frac{1}{V}\int_V u_r \rm{d}\Omega$, where $u_r = \lVert \tilde{\boldsymbol{u}}-\hat{\ub} \rVert / (\lVert \hat{\ub} \rVert + u_0)$ represent the relative displacement error between the PANN prediction $\tilde{\boldsymbol{u}}$ and the ground truth $\hat{\ub}$, and $u_0$ is set to $10^{-9}$mm to prevent the denominator from being zero.
}\label{fig:mean_disp_error}
\end{figure}

Simply comparing the absolute error cannot provide us with the accuracy of the method. 
Therefore, we calculated the relative displacement error at each FEM node and displayed the results together with the nodal displacement below corresponding absolute error contour plots in Fig.~\ref{fig:training_disp_error} and Fig.~\ref{fig:test_disp_error}.
The relative displacement error is defined as $u_r = u_e/ (\lVert \hat{\ub} \rVert + u_0)$. 
A small $u_0 = 10^{-9}$mm is used to make sure that the denominator does not become 0.
As we can observe from the figures, the displacement calculated by the PANN constitutive models closely align with the ground truth at all nodes.
The maximum relative errors are less than $0.8\%$ for the NH and MR models and less than $3\%$ for the AB model.
For the Gent and Fung models, the full-field average relative error are less than $1.5\%$, see Fig.~\ref{fig:mean_disp_error}.
Importantly, the nodes with large relative errors are primarily located in the nodes where displacement is small or near the fixed boundary. 
For example, for the training sample with Fung model, the node with maximum relative error has a displacement magnitude of 0.02 mm, which is much smaller than the displacement compared to other nodes. 
In those region, a rather small error can lead to a large relative error since the denominator is close to 0.
This highlighted the high accuracy of the trained PANN constitutive models in reproducing the displacement fields.

\subsection{Generalization to unseen loading}

For further validation on the extrapolation ability of the trained constitutive models, we deploy it for the calculation of four completely different problems.
We choose four different tests here, namely the uniaxial and biaxial tension/compression loading, shear and tension/compression combined with torsion. 
These tests are carefully chosen as they are rather common characterization techniques in experiments. 
For example, tension/compression combined with torsion can be realized using rheometers combined with some axial loading instruments, whereas unaxial tension is one of the most commonly used characterization techniques for hyperelastic materials. 
Specifically, the uniaxial loading, shear and biaxial loading tests were conducted on a $1~\text{mm} \times 1~\text{mm}$ square sample, while the tension/compression-torsion test was conducted on a cylinder sample with both radius and height of $1~\text{mm}$.
%Note that no data on compression was used during the training process.

\subsubsection{Uniaxial and shear loading}

\begin{figure}[!h]
\centering
\includegraphics[width=1.\textwidth]{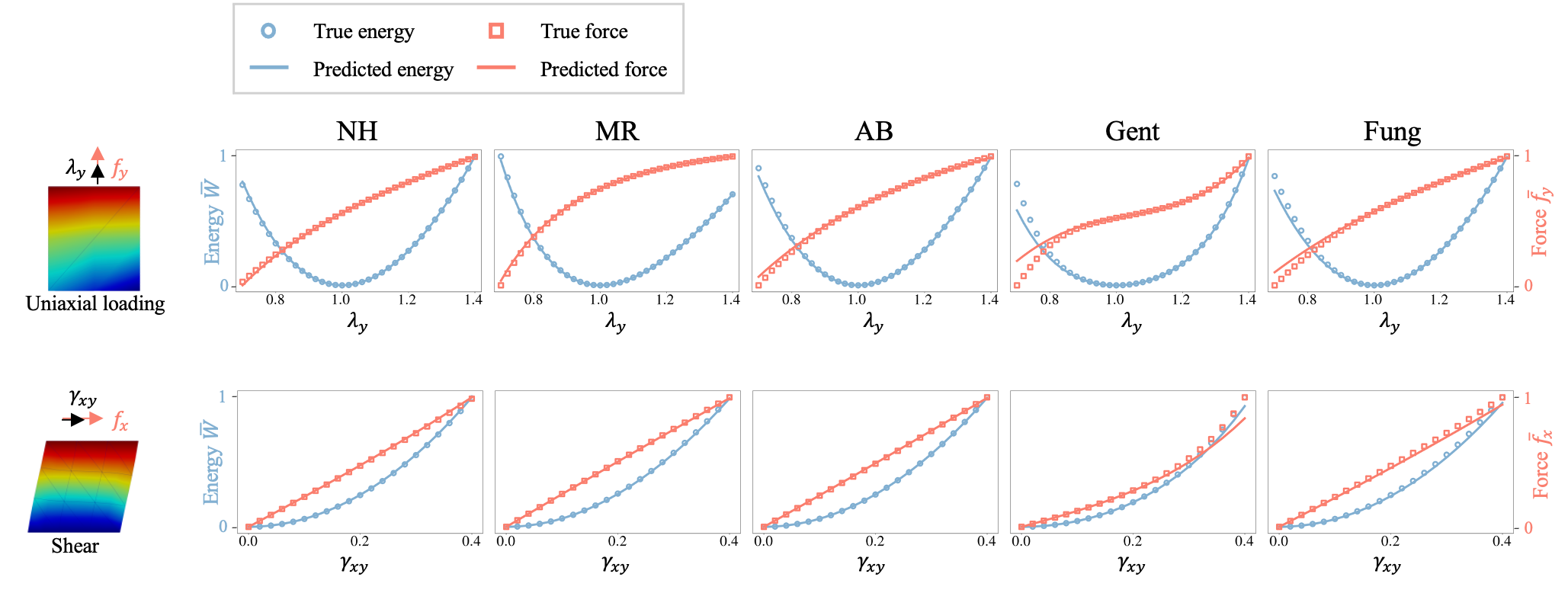}
\caption{Comparisons of the normalized strain energy densities and the axial/shear forces calculated by FEM with the ground truth and the trained NN-based constitutive models under the uniaxial and shear loading tests.
%The light purple regions in represent the principal stretch ranges extracted from the training data.
}\label{fig:uniaxial_loading_results}
\end{figure}

The comparisons of the normalized strain energy densities and axial/shear forces under the uniaxial/shear loading tests are illustrate in Figure.~\ref{fig:uniaxial_loading_results}.
In the test, we cover a wide range of loading conditions, with $\lambda_y \in [-0.7, 1.4]$ and $\gamma_{xy} in [0, 0.3]$. 
%The light purple regions in the uniaxial results represent the ranges of principal stretch in the training data.
Note that the data used in training is mainly generated by applying tensile loading to the sample up to a stretch of 20\%, and the portion over which the sample undergoes compressive deformation is highly limited. 
As we can easily observe that, the trained models exhibit high predictive accuracy for all five models within the stretching regions~($\lambda_y > 1$). 
Even in the compression regions~($\lambda_y < 1$), the PANN models can accurately capture the NH, MR, AB and Fung models.
And the performance of the trained model is less accurate for the Gent model, possibly due to its strong difference in the compression and tension behaviors.
In the shear test, the accuracy of the PANN models is high in all five models for both the elastic energy and the shear force.
%This shows that the PANN models have good predictability for a wide range of unseen loading conditions. 

\subsubsection{Biaxial loading}

\begin{figure}[!h]
\centering
\includegraphics[width=\textwidth]{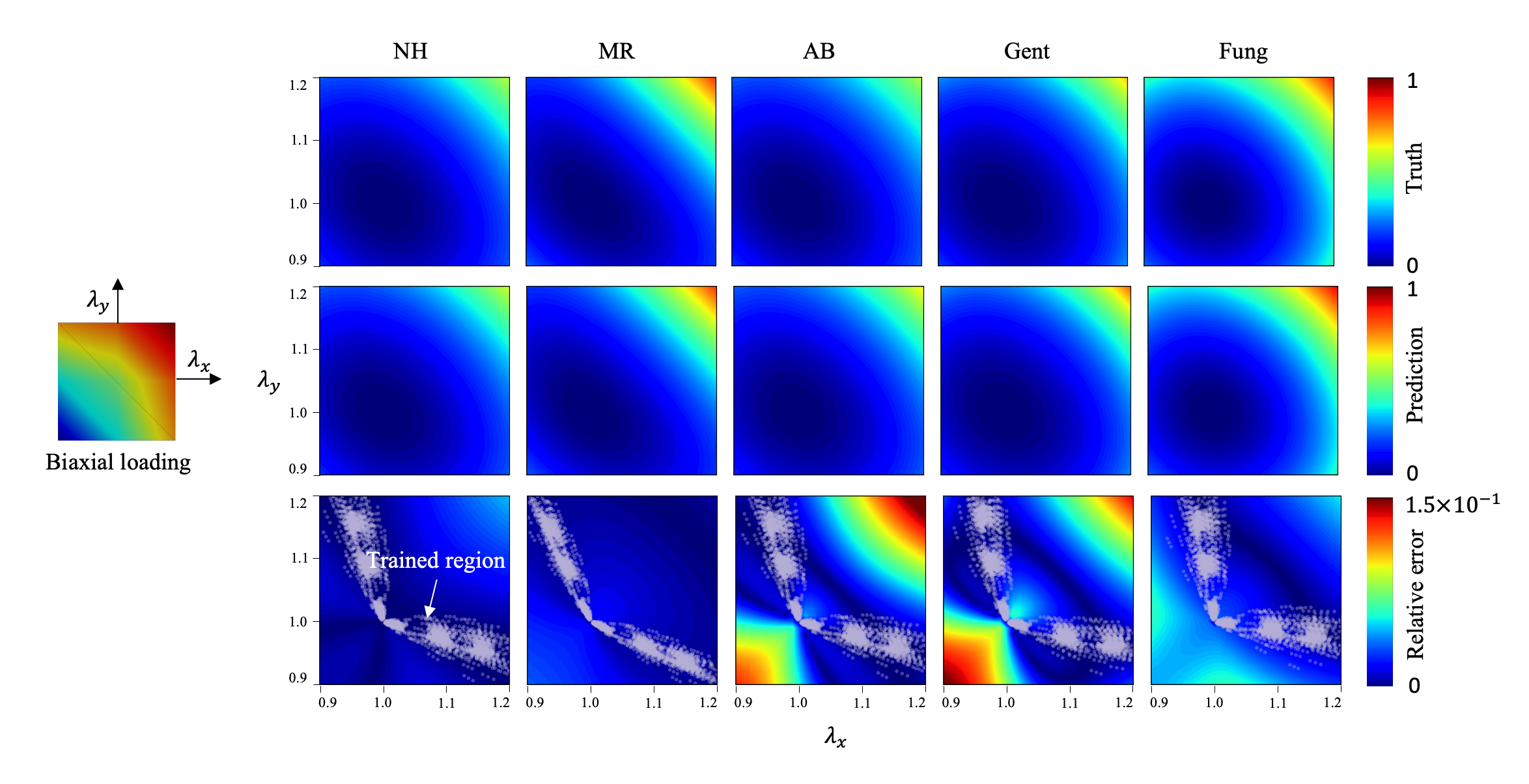}
\caption{Comparisons of the normalized strain energy densities calculated by FEM with the the ground truth and the trained NN-based models under the biaxial loading test.
The light purple circles in the error contours represent the trained deformation states extracted from the training data.
%The relative prediction error $W_e$ is defined as $W_e = |\widetilde{W} - \hat{W}|/|\hat{W} + 1\times 10^{-6}|$ where $\hat{W}$ and $\widetilde{W}$ are the ground truth and predicted strain energy densities respectively.
}\label{fig:biaxial_tension_test_results}
\end{figure}

The comparisons of the ground truth and the trained constitutive models under the biaxial loading are shown in Fig.~\ref{fig:biaxial_tension_test_results}.
The relative energy prediction error is defined as $W_e = |\widetilde{W} - \hat{W}|/|\hat{W} + 10^{-6}|$ where $\hat{W}$ and $\widetilde{W}$ represent the ground truth and predicted strain energy densities respectively.
The light purple circles in the error maps of Fig.~\ref{fig:biaxial_tension_test_results} are the principal stretches extracted from the training data, representing the deformation states covered in the dataset.
The distribution of these circles reveals that the training data mainly concentrates on the tensile deformation, which again explains why the trained models demonstrate better predictability in the stretching region, as shown in Fig.~\ref{fig:uniaxial_loading_results}.
The strain energy densities calculated through the trained models show strong agreement with the ground truth within the trained regions and across large surrounding areas.
Remarkably, although the error primarily aligns in the region that is far away from the light purple circles, the relative error in the NH, and MR models are still small~($<3\%$).
For Fung model, the maximum relative error is around $6\%$.
Such result further confirms that the trained models have not only good interpolation but also good extrapolation capability.

%\subsection{Shear}

%This could be attributed to the predominance of the stretching deformation mode in the training data~(see Fig.\ref{fig:biaxial_tension_test_results}(a)).
%The shear test results demonstrate in Fig.~\ref{fig:uniaxial_loading_results}(b) show the similar findings, i.e., the accuracy of the {MINN} models is high when $\gamma_{xy} < 0.3$ and gradually decreases as $\gamma_{xy} > 0.3$, and the trained {MINN} model failed to capture the energy divergence behavior of the Gent model with increasing load.
%These results could be explained as a transition form the trained region to the untrained region as $\gamma_{xy}$ increases.

\subsubsection{Torsion combined with tension/compression}

\begin{figure}[!h]
\centering
\includegraphics[width=\textwidth]{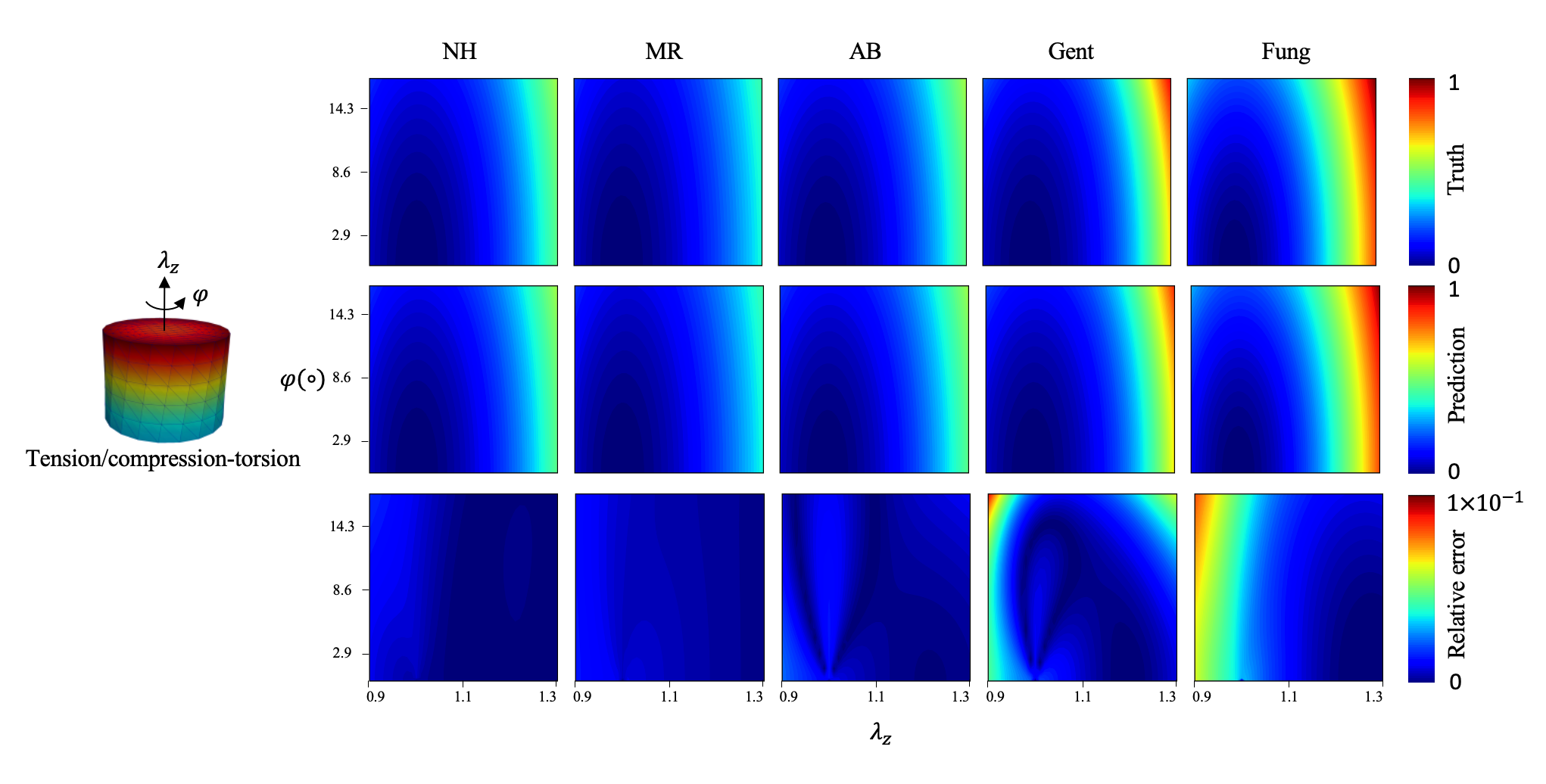}
\caption{Comparisons of the normalized strain energy densities calculated by FEM with the the ground truth and the trained NN-based models under the torsion combined with tension/compression loading test.
%The relative prediction error $W_e$ is defined as $W_e = |\widetilde{W} - \hat{W}|/|\hat{W} + 1\times 10^{-6}|$ where $\hat{W}$ and $\widetilde{W}$ are the ground truth and predicted strain energy densities respectively.
}\label{fig:torsion_test_results}
\end{figure}

Comparisons of the normalized strain energy densities calculated by FEM with the the ground truth and the trained models under the torsion combined with tension/compression loading test are shown in Fig.~\ref{fig:torsion_test_results}.
%The test was conducted on a cylinder sample with both radius and height of $1~\text{mm}$ in 3D geometry, and the loading was prescribed using a displacement boundary condition.
%The stress-strain state in such test is not included in the training data.
From the figure we can clearly observe that the relative error of the energy density $W_e$ is minimal in the tensile region~($\lambda_z > 1$) for all five models.
We in fact have covered a quite wide range of stretching and torsional angles, with the maximum stretch up to $30\%$, and a torsional angel up to 17.2$^{\circ}$.
The maximum relative errors within the tensile region are $0.9\%$, $0.9\%$, $1.4\%$, $6.2\%$, $4.4\%$ for the NH, MR, AB, Gent and Fung models respectively.
This clearly shows that the PANN models were able to learn the full 3D material behavior of the material based on the 2D data. 

Larger errors are found within the compressive region in Gent and Fung models. 
Note that even at a compression of 10\%, the maximum error of the models trained for NH, MR and AB models is less than 2\%, showing again that our optimization method can well capture the essence of these three models.   
For the Gent and Fung models, as in the previous cases, show less accuracy in the compressive region. 
But still, when the compression are small, the error is still rather small. 
For example, for a torsional angle of 17.2$^{\circ}$, under $5\%$ compression, the relative error for Gent and Fung models $4.5\%$ and $6.3\%$ respectively.

\section{Conclusion}
% {\bf Add some discussion on the generalizability of the work, i.e., how the parameters scale with the size of the neural networks, so that it can be used for more complex materials}
In summary, we have developed a PDE-constrained optimization method to efficiently learn hyperelastic constitutive relations from experimentally measurable full-field displacement and external load data obtained from a non-uniformly deformed sample. 
The constitutive relation is constructed on a modified physics-augmented neural networks~(PANN) to ensure that restrictions such as objectivity, balance of angular momentum, material symmetry, volumetric growth condition, polyconvexity and stress-free conditions are satisfied by construction.
Such architecture ensures that our NN-based constitutive model is physically consistent and our optimization algorithm mathematically well-posed.
%a {MINN} constitutive model and developed a hybrid FEM-NN optimization framework to efficiently calibrated the model on the experimentally measurable full-field displacement and external load data, enabling the discovery of the unknown hyperelastic constitutive models of real-world materials.
%physical constraints, including objectivity, material symmetry, polyconvexity, growth condition and stress-free condition are theoretically embedded to the {MINN} model.
%These constraints enable the implementation of the FEM simulation during the optimization process.
The performance of our method was tested on synthetic data obtained from numerical simulation.
Uniaxial tension test on a 2D plate with randomly placed circular holes is used to generate the training data, and due to the non-uniform deformation, one single test is enough for generating the training data. 
After the model is trained, we compare the prediction of the trained with the ground truth in geometries and complex loading conditions that are not included in the training data. 
Remarkably, the trained NN-based models can accurately capture the mechanics of the training materials even for completely unseen stress-strain states in the training data.
These testing results clearly shows that the NN-based model trained through the optimization method presented in this work has not only good interpolation, but also extrapolation capabilities.

It is important to note that, theoretically, the computational efficiency of the adjoint method-based optimization framework employed in this work will be hindered by an increasing number of optimization parameters.
This is because the complexity of solving both the finite element equations and the adjoint equations is independent of the number of optimization parameters.
This implies that the optimization framework can be efficiently applied to construct constitutive models for more complex material, such as anisotropic hyperelastic materials, fiber-reinforced materials, and fibrous networks, where larger neural networks are needed.
Our future work will focus on exploring the potential applications of the method to these material systems.
Expanding on this, generally speaking, a more diverse training data can broaden the NN's predictability. 
In this work, we have introduced randomly distributed circular holes in the specimens to generate non-uniform displacement training data. 
It is worth noting that in a recent work~\cite{tung2024anti}, studies have been conducted on maximizing the diversity of strain-stress states during experimental processes by designing the geometry of specimens.
Future work could explore integrating the optimization method with such efforts to further enhance the training efficiency of the NN-based constitutive models and expand its applicability range.

%Lastly, we expect that our approaches could further cement the use of NN-based constitutive models in materials science and computational mechanics.

\section*{Acknowledgment}
We acknowledge the partial support from the National Natural Science Foundation of China under the grant No. 12272005.
The work is also supported by the high-performance computing platform of Peking University

\bibliographystyle{elsarticle-num}

\bibliography{mybib}

\end{document}